\documentclass[12pt]{article}
\usepackage{epsfig,amsfonts,amssymb}
\usepackage{hyperref}
\usepackage{cite}
\input epsf.sty
\usepackage{cite}

\def\ZZZ{{\hbox{ Z\kern-1.6mm Z}}}
\def\RRR{{\hbox{ R\kern-2.4mm R}}}
\def\CCC{{\hbox{ C\kern-2.0mm C}}}
\def\zzz{{\hbox{z\kern-1mm z}}}

\newcommand{\qeq}{{\hbox{=\kern-2.3mm ? \kern.5mm }}}
\renewcommand{\qeq}{=}

\newcommand{\vp}{\varphi}

\newcommand{\VV}{{\cal V}}

\newcommand{\GG}{{\cal G}}

\newcommand{\HH}{{\cal H}}
\newcommand{\MM}{{\cal M}}

\newcommand{\OO}{{\cal O}}

\newcommand{\PP}{{\cal P}}

\newcommand{\XX}{{\cal X}}

\newcommand{\wt}{\widetilde}
\newcommand{\wh}{\widehat}

\newcommand{\RR}{{\cal R}}

\newcommand{\be}{\begin{equation}}
\newcommand{\ee}{\end{equation}}
\newcommand{\ben}{\begin{eqnarray}\displaystyle}
\newcommand{\een}{\end{eqnarray}}

\newcommand{\refb}[1]{(\ref{#1})}
\newcommand{\p}{\partial}
\newcommand{\sectiono}[1]{\section{#1}\setcounter{equation}{0}}

\def\one{{\hbox{ 1\kern-.8mm l}}}
\def\zero{{\hbox{ 0\kern-1.5mm 0}}}

\newcommand{\bea}[1]{\begin{eqnarray}\label{#1} }
\newcommand{\eea}{\end{eqnarray}}

\newcommand{\eqref}{\refb}

\usepackage{bm}
\usepackage[table]{xcolor}

\newcommand{\cL} {\{\hskip -4pt\{}
\newcommand{\cR} {\}\hskip -4pt\}}

\newcommand{\oR}{{\overline{\RR}}}

\begin{document}

\begin{flushright}
\end{flushright}

\vskip 12pt

\baselineskip 24pt

\begin{center}
{\Large \bf  Reality of Superstring Field Theory Action}

\end{center}

\vskip .6cm
\medskip

\vspace*{4.0ex}

\baselineskip=18pt

\centerline{\large \rm Ashoke Sen}

\vspace*{4.0ex}

\centerline{\large \it Harish-Chandra Research Institute}
\centerline{\large \it  Chhatnag Road, Jhusi,
Allahabad 211019, India}

\vspace*{1.0ex}
\centerline{\small E-mail:  sen@mri.ernet.in}

\vspace*{5.0ex}

\centerline{\bf Abstract} \bigskip

We determine the reality conditions on the string fields that make the action for
heterotic and type II string field theories real.

\vfill \eject

\baselineskip 18pt

\tableofcontents

\sectiono{Introduction and summary} \label{sintro}

String field theory is a useful technique that allows us to systematically deal
with the infrared divergences that arise in the usual world-sheet approach.
For this program to be successful, one needs to construct a suitable string field theory
whose Feynman diagrams reproduce the string theory amplitudes constructed from
world-sheet description, to all orders in perturbation expansion. Such an action
was constructed for bosonic string theories in \cite{wittensft,9206084}, 
and for type II and heterotic
string theories more recently in \cite{1508.05387}.

Given a string field theory action, one important question is: what are the reality
conditions on the fields that appear in the action? These conditions ensure that
the action is real for arbitrary field configuration satisfying the reality conditions.
One can carry out much of the analysis in the theory, {\it e.g.} proof of gauge invariance,
derivation of Feynman rules, etc.\ without knowing the reality conditions. Nevertheless
being able to find this condition is necessary for the consistency of the theory. 
For example this is necessary for
determining the overall phase of the S-matrix, which in turn is needed for checking
the  unitarity of the S-matrix\cite{1604.01783}. It is also necessary for determining
which classical solutions are allowed. For example if we have a scalar field
with potential proportional to $(\phi^2+a^2)^2$, then the only translationally invariant
solution is $\phi=0$ if $\phi$ is required to take real values, whereas we can also have
solutions with $\phi=\pm i\, a$ if $\phi$ is required to take imaginary values.

Reality conditions for the fields of bosonic string theory were determined in
\cite{9206084} and analyzed in more detail in \cite{9705038}. 
In this paper we determine the reality conditions for the fields of
superstring field theory. Our method differs slightly from that of \cite{9206084}.
So in \S\ref{sboson} we first illustrate this method by applying it to the
bosonic string field theory. The result of our analysis agrees with that of \cite{9206084}.
In \S\ref{sfermion} we apply this method to determine the reality condition on
the fields of superstring field theory constructed in \cite{1508.05387}. 
In \S\ref{srelation} we rewrite the reality condition as a relation between hermitian
conjugate and BPZ conjugate of the string field, generalizing the result of
\cite{9206084}.
In \S\ref{sgen} we
briefly discuss extension of our analysis to bosonic and superstring field theories
in arbitrary background described by general world-sheet (super-)conformal field
theory.

Throughout this paper we shall follow the conventions of \cite{1508.05387}. 
These differ from
those of \cite{9206084} in certain aspects. 
For example the bracket $\cL~\cR$ used here
was denoted simply by $\{~\}$ in \cite{9206084} 
and the regions $\oR_{g,n}$ were called 
$\VV_{g,n}$ in \cite{9206084}. 
Our normalization condition for the correlation functions
in the world-sheet theory is given in \refb{enorm} which differs
from the one used in \cite{9206084} by a minus sign.

\sectiono{Reality condition in bosonic string field theory} \label{sboson}

The world-sheet theory of bosonic string theory contains 26 scalars
$X^\mu$ for $0\le\mu\le 25$, 
holomorphic ghost fields $b,c$, and anti-holomorphic ghost fields
$\bar b,\bar c$. The singular parts of the operator product
expansion of these fields have the form:
\ben
&& b(z) c(w) = {1\over z-w}+\cdots, \quad \bar b(\bar z)\bar c(\bar w) =
{1\over \bar z -\bar w}+\cdots, \nonumber \\
&& \p X^\mu(z) \p X^\nu(w) = -{\eta^{\mu\nu}\over 2 (z-w)^2}+\cdots,
\quad \bar\p X^\mu(\bar z) \bar\p X^\nu(\bar w) = -{\eta^{\mu\nu}
\over 2 (\bar z-\bar w)^2}+\cdots\, ,
\een
where we have set $\alpha'=1$.
On the complex plane, the fields have mode expansion
\ben
&& b(z) = \sum_n b_n z^{-n-2}, \quad c(z) = \sum_n c_n z^{-n+1}, \quad 
\bar b(\bar z) = \sum_n \bar b_n \bar z^{-n-2}, \quad \bar c(\bar z) = \sum_n 
\bar c_n \bar z^{-n+1}\, , \nonumber \\
&& i \, \sqrt 2\, \p X^\mu(z) = \sum_n \alpha^\mu_n z^{-n-1}, \quad i \sqrt 2\,
\bar \p X^\mu(\bar 
z) = \sum_n \bar\alpha^\mu_n \bar z^{-n-1}, \quad \bar\alpha^\mu_0\equiv \alpha^\mu_0\, .
\een
The Virasoro generators, defined through the mode expansion of the
stress tensors $T(z)$ and $\bar T(\bar z)$:
\be
T(z)=\sum_n L_n z^{-n-2}, \quad \bar T(\bar z) =\sum_n \, \bar L_n \bar z^{-n-2}\, ,
\ee
can be expressed in terms of $b_n$, $\bar b_n$, $c_n$, $\bar c_n$, $\alpha^\mu_n$ and
$\bar\alpha^\mu_n$ without any explicit factor of $i$. 
We shall denote by $\HH$ the Hilbert space of states $|s\rangle$ in the
combined CFT of the matter and ghost system satisfying
\be \label{el0}
|s\rangle \in \HH \quad \hbox{iff} \quad b_0^-|s\rangle=0, \quad L_0^-|s\rangle=0\, ,
\ee
where
\be
b_0^\pm \equiv (b_0\pm \bar b_0), \quad L_0^\pm \equiv (L_0\pm \bar L_0),
\quad c_0^\pm \equiv {1\over 2} (c_0\pm \bar c_0)\, .
\ee
Let
$\{|\vp_r(k)\rangle\}$ be a complete set of basis states created by the action
of $b_{-n}$, $\bar b_{-n}$ for $n\ge 2$, 
$c_{-n}$, $\bar c_{-n}$ for $n\ge -1$, and $i\alpha^\mu_{-n}$,
$i\bar\alpha^\mu_{-n}$ for $n\ge 1$
on the  state $|k\rangle=e^{ik\cdot X}(0)|0\rangle$, where
$|0\rangle$ denotes the SL(2,C) invariant vacuum. In that case
the vertex operators of the states $|\vp_r\rangle$ can be expressed as sum of
products
of $e^{ik\cdot X}$ and (derivatives of) $b$, $c$, $\bar b$, $\bar c$, $\p X$ and
$\bar\p X$, without any explicit factor of $i$. 

The string field $|\Psi\rangle$ is taken to be an arbitrary state in $\HH$, and can 
be expanded as
\be \label{eexpand}
|\Psi\rangle = \sum_r \int {d^{26} k\over (2\pi)^{26}} \, \psi_r(k) |\vp_r(k)\rangle\, .
\ee
If $\vp_r$ has ghost number $n_r$, then $\psi_r$ and $\vp_r$ have grassmann
parity $(-1)^{n_r}$ so that the string field is always grassmann even.
The string field theory action has a kinetic term
\be\label{eactfree}
S_K = {1\over 2g_s^2} \langle\Psi|c_0^- Q_B |\Psi\rangle =
{1\over 2 g_s^2} \sum_{r,s} \, (-1)^{n_r n_s}
\int {d^{26} k\over (2\pi)^{26}} {d^{26} k'\over
(2\pi)^{26}} \, 
\psi_r(k) \psi_s (k') \langle \vp_r(k) | c_0^- Q_B |\vp_s(k')\rangle
\ee
where $g_s$ is the string coupling and 
$Q_B$ is the BRST charge, constructed from the oscillators of $b$, $c$,
$\bar b$, $\bar c$, $T$ and $\bar T$ without any explicit factor of $i$.
$\langle A|B\rangle$ denotes BPZ inner product defined
 as
\be \label{ebpz}
\langle A|B\rangle = \langle I\circ A(0) B(0)\rangle\, ,
\ee
with $I(z)\equiv 1/z$ and $I\circ A$ denoting the conformal transform of $A$ by
the transformation $I$.

In order to construct the interaction term, we introduce a fiber bundle
$\wh\PP_{g,n}$ with base $\MM_{g,n}$ -- the moduli space of genus
$g$ Riemann surface with $n$ punctures -- and fiber labelled by the choice of local
coordinates (up to phases) around each puncture\cite{9206084}.
We shall denote by
$\Sigma_{g,n}$ a point in $\wh\PP_{g,n}$ describing a specific Riemann surface with
$n$-punctures and the choice of local coordinates on the punctures.
The multi-string
interaction vertex $\cL A_1\cdots A_n\cR$ 
for
arbitrary states $|A_1\rangle,\cdots |A_n\rangle\in \HH$
in then defined as
\be\label{edefcurly}
\cL A_1\cdots A_n\cR =  \sum_g (g_s)^{2g} \, (2\pi i)^{-(3g-3+n)} \, 
\int_{\oR_{g,n}} d m^1\wedge \cdots \wedge dm^{6g-6+2n} \, 
\left\langle b[v^{(1)}]\cdots b[v^{(6g-6+2n)}] \prod_{i=1}^n 
A_i\right\rangle_{\Sigma_{g,n}}\, ,
\ee
where
$\oR_{g,n}$ denotes part of a section of $\wh\PP_{g,n}$ satisfying appropriate
identities\cite{9206084},
$m^1,\cdots m^{6g-6+2n}$ are the coordinates on $\MM_{g,n}$ which can also
be taken to parametrize $\oR_{g,n}$, 
and $\langle ~\rangle_{\Sigma_{g,n}}$
denotes correlation function on the Riemann surface $\Sigma_{g,n}$, with the
vertex operators of $A_1,\cdots A_n$ inserted at the punctures using the local
coordinate system associated with $\Sigma_{g,n}$. The $b[v^{(\alpha)}]$ factors are
defined as follows. We can use standard procedure involving
Schiffer variation\cite{9206084} to associate with
any tangent vector $\p/\p m^\alpha$ of $\oR_{g,n}$ a set of holomorphic
vector fields 
$v^{(\alpha,i)}$  on $\Sigma_{g,n}$ for $i=1,\cdots n$. $v^{(\alpha,i)}$ 
either vanishes or 
is well defined around the curve $C_i$ encircling the $i$-th
puncture, but may not be well defined away from $C_i$.
Then $v^{(\alpha,i)}$ will have a Laurent expansion
of the form
\be 
v^{(\alpha,i)}(w_i) = \sum_m v^{(\alpha,i)}_m w_i^{-m+1}\, ,
\ee
where $w_i$ denotes the local coordinate around the $i$-th puncture with
the puncture situated at $w_i=0$.
In this case we define
\be
b[v^{(\alpha)}] = \sum_i\left[\ointop_{C_i} dw_i\, b(w_i) \, v^{(\alpha,i)}(w_i) 
+ \ointop_{C_i} d\bar w_i\, \bar b(\bar w_i) \, \overline{v^{(\alpha,i)}(w_i)}
\right]\, ,
\ee
where the definition of $\ointop$ includes the usual $1/2\pi i$ factors
so that $\ointop_{C_i} dw_i/w_i=1$ and $\ointop_{C_i} d\bar w_i/\bar w_i=1$. 
If $b^{(i)}_n$
and $\bar b^{(i)}_n$ denote the usual oscillators of $b$ and $\bar b$ acting on the
Hilbert space of the $i$-th external state, then this can also be expressed as
\be \label{ebv}
b[v^{(\alpha)}] = \sum_i \sum_m \left(v^{(\alpha,i)}_m b^{(i)}_{-m} 
+ v^{(\alpha,i)*}_m
\bar b^{(i)}_{-m} \right)\, ,
\ee
where $*$ denotes complex conjugation.

In terms of the curly bracket defined in \refb{edefcurly}, the interaction term of the
string field theory action takes the form
\be \label{eactint}
S_I = {1\over g_s^2} \sum_{n=1}^\infty {1\over n!} \, \cL \Psi^n\cR\, ,
\ee
where $\cL\Psi^n\cR\equiv \cL\Psi \Psi \cdots \Psi\cR$ with $n$ insertions of $\Psi$ 
inside the curly bracket. Note that the sum starts at $n=1$. While the tree level action
contains interaction terms involving cubic and higher powers of the string field,
the Batalin-Vilkovisky (BV) quantum
master action $S_K+S_I$ also receives higher genus contribution that includes
linear and quadratic terms in the string field.

Let us denote by $\Sigma^*_{g,n}$ the
Riemann surface of genus $g$ and $n$ punctures, obtained from
$\Sigma_{g,n}$ by complex conjugation of all transition functions used to define 
$\Sigma_{g,n}$, and the local coordinates at the punctures. We also denote
by
$\oR^*_{g,n}$ the image of $\oR_{g,n}$ under this map. 
We shall assume that $\oR_{g,n}$ has been chosen such 
that\cite{9206084}\footnote{Since at a generic point on the moduli space of
Riemann surfaces the conjugation acts non-trivially, \refb{econdi} requires
that on the conjugate Riemann surface we choose the local coordinates to be
complex conjugates of the original choice. On special Riemann surfaces which
are invariant under conjugation, \refb{econdi} requires that the local
coordinates must either be invariant under conjugation, or we must average
over two choices related by conjugation.}
\be\label{econdi}
\oR^*_{g,n} = \oR_{g,n}\, .
\ee
This means that for every $\Sigma_{g,n}\in\oR_{g,n}$, we have $\Sigma^*_{g,n}\in 
\oR_{g,n}$. 

We are now ready to describe the reality condition on the string field
$|\Psi\rangle$. 
If $n_r$ is the ghost 
number of $\vp_r$, then we impose the reality condition
\be \label{erealpsi}
\psi_r(k)^* = (-1)^{n_r (n_r+1) /2 + 1}\psi_r(-k)\, ,
\ee
where $\psi_r(k)$ are the coefficients of expansion appearing in 
\refb{eexpand}.
Therefore $\psi_r(k)^*=\psi_r(-k)$ when $n_r=1$ or $2$ mod 4, 
and $\psi_r(k)^*=-\psi_r(-k)$ when $n_r=0$ or
$3$ mod 4. 
Our goal will be to show that once $\psi_r(k)$ satisfy \refb{erealpsi},
the string field theory action given by the sum of
\refb{eactfree} and \refb{eactint}
takes real values.

If we define $\chi_r(k)$ via
\be\label{ephase}
\psi_r(k) = (i)^{n_r (n_r+1) /2 + 1}\chi_r(k)\, ,
\ee
then the reality condition may be written as
\be\label{echireal}
\chi_r(k)^* = \chi_r(-k)\, .
\ee
Alternatively we could have absorbed the phase factor on the right hand
side of \refb{ephase} into the definition of the basis states $|\vp_r(k)\rangle$
so that $\chi_r(k)$ will be directly the coefficients appearing in the expansion
of the string field in this basis. 
However we shall continue to work with the original choice of the basis states.

In terms of the variables $\chi_r(k)$ the
bosonic string field theory action given by the sum of \refb{eactfree} and
\refb{eactint} may be written as
\be \label{eaction}
S=g_s^{-2} \sum_n {1\over n!} \sum_{r_1,\cdots r_n}
(-1)^{\sum_{i<j} n_{r_i} n_{r_j}}
\int {d^{26} k_1\over (2\pi)^{26}}\cdots {d^{26} k_n \over (2\pi)^{26}}\,
V^{(n)}_{r_1\cdots r_n}(k_1,\cdots k_n) \, \chi_{r_1}(k_1)\cdots \chi_{r_n}(k_n)
\, ,
\ee
where the vertex $V^{(n)}$ is given by
\ben \label{econs}
V^{(n)}_{r_1\cdots r_n}(k_1,\cdots k_n) 
&=& (i)^{\sum_{i=1}^n \{n_{r_i} (n_{r_i}+1)/2 + 1\}} \cL\vp_{r_1}(k_1) \cdots \vp_{r_n}(k_n)\cR\, ,
\quad \hbox{for} \quad n\ne 2, \nonumber \\
V^{(2)}_{r_1 r_2}(k_1, k_2) 
&=& (i)^{\sum_{i=1}^2 \{n_{r_i} (n_{r_i}+1)/2 + 1\}} \left[
\langle \vp_{r_1}(k_1)|c_0^- Q_B 
|\vp_{r_2}(k_2)\rangle
+\cL\vp_{r_1}(k_1) \vp_{r_2}(k_2)\cR\right]\, , \nonumber \\
\een
with the `interaction terms' $\cL\vp_{r_1}(k_1) \cdots \vp_{r_n}(k_n)\cR$ for $n\le 2$
receiving contributions from genus $\ge 1$ Riemann surfaces.
This has  symmetry property:
\be \label{eexchange}
V^{(n)}\to (-1)^{n_{r_i} n_{r_{i+1}}} V^{(n)} \quad \hbox{under} \quad r_i\leftrightarrow r_{i+1},
\quad k_i\leftrightarrow k_{i+1}\, ,
\ee
and satisfies ghost number and momentum conservation laws
\be \label{econ}
V^{(n)}_{r_1\cdots r_n}(k_1,\cdots k_n)  \propto 
\delta_{\sum_i n_{r_i}, 2n} \,  \delta^{({26})}(k_1+\cdots k_n)\, .
\ee
We shall show that 
\be \label{evpreal}
\cL\vp_{r_1}(k_1) \cdots \vp_{r_n}(k_n)\cR^* 
= \cL\vp_{r_1}(-k_1) \cdots \vp_{r_n}(-k_n)\cR\, ,
\ee
and
\be \label{evpreal2}
\langle \vp_{r_1}(k_1)|c_0^- Q_B 
|\vp_{r_2}(k_2)\rangle^* = \langle \vp_{r_1}(-k_1)|c_0^- Q_B 
|\vp_{r_2}(-k_2)\rangle\, .
\ee
It follows from this that
\be 
V^{(n)}_{r_1\cdots r_n}(k_1,\cdots k_n)^*
= 
(-1)^{\sum_{i=1}^n \{n_{r_i} (n_{r_i}+1)/2 + 1\}} V^{(n)}_{r_1\cdots r_n}
(-k_1,\cdots -k_n)\, .
\ee
Using \refb{eexchange} this can be written as
\ben
V^{(n)}_{r_1\cdots r_n}(k_1,\cdots k_n)^*
&=& (-1)^{\sum_i \{n_{r_i} (n_{r_i}+1)/2 + 1\}}  (-1)^{\sum_{i<j} n_{r_i} n_{r_j}}
V^{(n)}_{r_n\cdots r_1}(-k_n,\cdots -k_1) \nonumber \\
&=& (-1)^{\sum_i \{n_{r_i}^2 + n_{r_i}+2\}/2} (-1)^{\sum_i n_{r_i} \sum_j n_{r_j} / 2 
- \sum_i n_{r_i}^2/2}
V^{(n)}_{r_n\cdots r_1}(-k_n,\cdots -k_1) \, . \nonumber \\
\een
Using the constraint on $n_{r_i}$ given in \refb{econ}, this can be
rewritten as
\ben
V^{(n)}_{r_1\cdots r_n}(k_1,\cdots k_n)^*
&=& (-1)^{\sum_i n_{r_i}^2 / 2 + n + n} (-1)^{2 n^2
- \sum_i n_{r_i}^2/2}
V^{(n)}_{r_n\cdots r_1}(-k_n,\cdots -k_1)  \nonumber \\
&=& V^{(n)}_{r_n\cdots r_1}(-k_n,\cdots -k_1) \, .
\een
Reality of the action \refb{eaction} follows immediately from this,
\refb{echireal}, and the fact
that under complex conjugation, a product of fields gets transformed to
the product of the complex conjugate fields in the reverse order. 
For grassmann even fields this order reversal has no effect, but for
grassmann odd fields this is related to the product in the original 
order by a sign.

It remains to prove \refb{evpreal} and \refb{evpreal2}. To prove \refb{evpreal2} we
note that after expressing $Q_B$
and the states $\vp_{r_1}(k_1)$ and $\vp_{r_2}(k_2)$   in terms of the matter and ghost oscillators, the only explicit factors of $i$ arise from the fact that in the expressions
for the states $\vp_r$'s we use the combination $-i\alpha_{-n}^\mu$ and 
$-i\bar\alpha_{-n}^\mu$. Now since the amplitude is Lorentz invariant, the 
$\alpha^\mu$'s must contract with each other in which case the factors of $i$ combine
in pairs to give a real number, or the $i\alpha^\mu_0$ factor acts on the vacuum producing 
a factor proportional to $i k^\mu$. Since the latter factor remains invariant under the
combined operation of complex conjugation and change of the sign of momenta,
we get \refb{evpreal2}.

The proof of \refb{evpreal} can be given as follows.
The general form of $\cL\vp_{r_1}(k_1)\cdots \vp_{r_n}(k_n)\cR$ is given by
\be \label{emult}
\sum_{g=0}^\infty (g_s)^{2g}\, (2\pi i)^{-(3g-3+n)} \, 
\int_{\oR_{g,n}} d m^1\wedge \cdots \wedge dm^{6g-6+2n} \, 
\left\langle b(v^{(1)})\cdots b(v^{(6g-6+2n)}) \prod_{i=1}^n 
\vp_{r_i}(k_i)\right\rangle_{\Sigma_{g,n}}\, .
\ee
First let us ignore the $(2\pi i)^{-(3g-3+n)}$ factor and the insertions of $b(v^{(i)})$'s
in \refb{emult}. 
In this case the correlation function 
\be
\left\langle\prod_{i=1}^n 
\vp_{r_i}(k_i)\right\rangle_{\Sigma_{g,n}} 
\ee
involves vertex operators constructed out of products of 
$b$, $c$, $\bar b$, $\bar c$, $\p X$, $\bar \p X$ and $e^{ik\cdot X}$ and their
derivatives. Since the operator products of these operators have no explicit factor
of $i$ except for the factor of $i$ accompanying each momentum factor $k^\mu$,
complex conjugation of the amplitude will have the effect of 
changing the sign of all the momentum
factors, and
mapping 
$\Sigma_{g,n}$ to $\Sigma^*_{g,n}$.\footnote{One way to see this is to
regard the genus $g$ Riemann surface with $n$ punctures
as the result of plumbing fixture of
several 3-punctured spheres. This allows us to express the correlation function on the
genus $g$ surface
in terms of products of three point functions on the sphere.
Let us for definiteness take the three insertion points on each
sphere to be on the real axis, {\it e.g.}
at 0, 1 and 2, and use the global coordinate on the complex plane in the plumbing
fixture relations, {\it e.g.} for gluing the puncture at 0 on the $i$-th sphere
to the puncture at 1 on the $j$-th sphere, use $z_i (z_j-1)=q_{ij}$.
Now since all the three point functions are real in the basis we have chosen 
-- except for the factors of $i$ multiplying $k^\mu$ -- complex conjugation of the
amplitude will have the effect of changing $k^\mu$ to $-k^\mu$, and complex
conjugating all the variables $\{q_{ij}\}$ 
appearing in the plumbing fixture relations. The latter
precisely takes us from $\Sigma_{g,n}$ to $\Sigma^*_{g,n}$.}
Therefore we get
\be \label{esigstar}
\left(\left\langle\prod_{i=1}^n 
\vp_{r_i}(k_i)\right\rangle_{\Sigma_{g,n}} \right)^*
= \left\langle\prod_{i=1}^n 
\vp_{r_i}(-k_i)\right\rangle_{\Sigma^*_{g,n}} 
\ee

Let us now study the effect of inserting the $b(v^{(\alpha)})$'s and the overall multiplicative
factor of $(2\pi i)^{-(3g-3+n)}$ as given in \refb{emult}. Complex conjugation
of the multiplicative factor gives a factor of $(-1)^{3g-3+n}$.
From \refb{ebv}
we see that
we can represent the 
effect of the $b(v^{(\alpha)})$ insertions by changing the tensor product of external states
$ |\vp_{r_1}(k_1)\rangle_{(1)} \otimes \cdots \otimes |\vp_{r_n}(k_n)\rangle_{(n)}$ to
\be
\prod_{\alpha=1}^{6g-6+2n} \left\{\sum_{i=1}^n \sum_{m=-\infty}^\infty 
\left(v^{(\alpha,i)}_{-m} b^{(i)}_{m} + v^{(\alpha,i)*}_{-m} \bar b^{(i)}_{m}\right)\right\}
 |\vp_{r_1}(k_1)\rangle_{(1)} \otimes \cdots \otimes |\vp_{r_n}(k_n)\rangle_{(n)}\, .
\ee
The action of $b_{m}$ and/or $\bar b_{m}$ on the $i$-th state is to change the
vertex operator to a different one that is still made of $b$, $c$, $\bar b$, $\bar c$,
$\p X$, $\bar \p X$, $e^{ik\cdot X}$ and their derivatives without any explicit
factor of $i$, except for the factors of $i$ accompanying each factor of $k_i^\mu$. 
Therefore the effect of complex conjugation of this amplitude will
be to evaluate the correlation function on 
$\Sigma^*_{g,n}$, change the sign of all
momenta, and replace $v^{(j,i)}_{-m}$ by its complex conjugate
$\left(v^{(j,i)}_{-m}\right)^*$.
We shall now compare 
$\left(v^{(j,i)}_{-m}\right)^*$'s with the $v^{(j,i)}_{-m}$'s associated with the
tangent vectors of $\oR_{g,n}$ around the point
$\Sigma^*_{g,n}$. 
There are several transformations involved in relating 
$v^{(j,i)}_{-m}$'s around the point
$\Sigma^*_{g,n}$ to $v^{(j,i)}_{-m}$'s around the point
$\Sigma_{g,n}$:
\begin{enumerate}
\item First of all since $\Sigma_{g,n}^*$ is obtained from $\Sigma_{g,n}$ by
complex conjugating all transition functions, 
the $v^{(j,i)}_{-n}$'s will get complex conjugated:
\be
v^{(j,i)}_{-n} \to v^{(j,i)*}_{-n}\, .
\ee
\item Since complex conjugation of the transition functions
used in defining the Riemann surface
induces complex conjugation of the coordinates on $\MM_{g,n}$, 
the integration measure $\prod_i
d\mu_i\wedge d\bar \mu_i$, where $\{\mu_i\}$ are the
complex moduli, transforms to $\prod_i d\bar \mu_i\wedge d \mu_i= 
\prod_i(- d\mu_i\wedge d\bar \mu_i)$.
Therefore
the orientation of
the moduli space picks a minus sign for each complex dimension. 
As a result, 
half of the $6g-6+2n$
$v^{(j,i)}_{-n}$'s also change sign besides being complex conjugated
when we compare the tangent vectors of $\oR_{g,n}$ 
around $\Sigma_{g,n}$ and
$\Sigma^*_{g,n}$.
This gives
a factor of 
\be \label{esgb}
(-1)^{3g-3+n}
\, ,
\ee
when we compare the integration measure around
$\Sigma_{g,n}$ with the integration measure around $\Sigma^*_{g,n}$.
\end{enumerate}

These two effects together lead to the equation
\ben \label{einter}
&& \left[(2\pi i)^{-(3g-3+n)} \, 
\int_{\oR_{g,n}} d m^1\wedge \cdots \wedge dm^{6g-6+2n} \, 
\left\langle b(v^{(1)})\cdots b(v^{(6g-6+2n)}) \prod_{i=1}^n 
\vp_{r_i}(k_i)\right\rangle_{\Sigma_{g,n}}\right]^*\nonumber \\
&=& (2\pi i)^{-(3g-3+n)} \,  
\int_{\oR^*_{g,n}} d m^1\wedge \cdots \wedge dm^{6g-6+2n} \, 
\left\langle b(v^{(1)})\cdots b(v^{(6g-6+2n)}) \prod_{i=1}^n 
\vp_{r_i}(-k_i)\right\rangle_{\Sigma_{g,n}}\, .\nonumber \\
\een
Note that the $(-1)^{3g-3+n}$  given in \refb{esgb} cancels the minus sign that 
arises from the complex conjugation of the $(2\pi i)^{-(3g-3+n)}$ factor.
On the right hand side we have replaced the subscript $\Sigma^*_{g,n}$ of
\refb{esigstar} by $\Sigma_{g,n}$ since conjugation operation is already encoded
in the fact that the integration is performed over $\oR^*_{g,n}$.
Using \refb{einter} and the fact that $\oR_{g,n}=\oR^*_{g,n}$, we recover
\refb{evpreal}.

This completes the proof of reality of the action of bosonic string field
theory.
One point worth mentioning here is that the reality conditions
on all the fields are not unambiguously fixed by demanding the reality of the
action. For example since for an $m$-point amplitude of states carrying
ghost numbers $n_1,\cdots n_m$ we have $\sum_k n_k = 2m$, the action remains
invariant if we scale the fields as
\be \label{eamb1}
\psi_r(k) \to e^{i\alpha (n_r-2)} \psi_r(k)\, ,
\ee
where $\alpha$ is an arbitrary real number.  Therefore
whatever reality condition was imposed on $\psi_r(k)$ can instead be imposed
on $e^{i\alpha (n_r-2)} \psi_r(k)$ without affecting the reality of the action.
Since this does not transform states
in the physical sector (which have $n_r=2$) this scaling has no effect on the
physical S-matrix of the theory.

Once the reality condition on the string field is determined, we can use this to fix
the overall sign of the action. 
Consider for example the string field component labelling the tachyon field 
$T(k)$
\be
\int {d^{26} k\over(2\pi)^{26}} \, T(k) \, \bar c \, c \,  e^{ik\cdot X}\, .
\ee
According to the reality condition \refb{erealpsi}, 
$T(k)$ is the Fourier transform of a real
scalar field. Using the normalization condition\footnote{This differs from that
of \cite{9206084} by a minus sign. The consequence of this different sign convention has
been discussed in \cite{1508.02481}.}
\be \label{enorm}
\langle k| c_{-1} \bar c_{-1} c_0 \bar c_0 c_1 \bar c_1|k'\rangle = 
(2\pi)^{26} \delta^{({26})}(k+k')\, ,
\ee
the kinetic term \refb{eactfree} includes a term
\be
 {1\over 4 g_s^2} \int {d^{26} k\over (2\pi)^{26}} \, \left({1\over 2} k^2 - 2\right) T(k) T(-k), \quad k^2\equiv -(k^0)^2 +\vec k^2\, .
\ee
This is a wrong sign for the kinetic term. This can be repaired by the substitution 
$g_s^2 \to -g_s^2$. As long as this substitution is made 
in the interaction term as well, the action
satisfies the requirement of gauge invariance, and we get a consistent 
string field theory. Furthermore, since the expansion of the action is in powers of
$g_s^2$, this does not introduce any extra factors of $i$, and the action remains 
real.

Before concluding this section we shall discuss the relation between the reality
condition \refb{erealpsi} and the one discussed in  \cite{9206084}. The reality condition
in  \cite{9206084} was stated as the requirement that the hermitian conjugate
and BPZ conjugate of a string field should have opposite signs. 
Therefore in order to translate this condition to a condition on the coefficients
$\psi_r(k)$ in \refb{eexpand}, we need to understand the difference between the action of
hermitian conjugation and BPZ conjugation:
\begin{enumerate}
\item Hermitian conjugation replaces the ket state $|k\rangle$ by the bra
$\langle -k|$, while BPZ conjugation replaces $|k\rangle$ by 
$\langle k|$.
\item Hermitian conjugation complex conjugates the coefficients $\psi_r(k)$
while BPZ conjugation leaves them unchanged.
\item Both hermitian conjugation and BPZ conjugation act in the same way
on the oscillators $b_n$, $\bar b_n$, $i\alpha^\mu_n$ and $i\bar\alpha^\mu_n$,
replacing $n$ by $-n$ and also changing the signs of $i\alpha^\mu_n$ and
$i\bar\alpha^\mu_n$. On the other hand hermitian conjugation takes
$c_n$ and $\bar c_n$ to $c_{-n}$ and $\bar c_{-n}$, while BPZ conjugation
takes them to $-c_{-n}$ and $-\bar c_{-n}$, respectively. In arriving at these signs we
have used the convention that BPZ conjugation involves conformal transformation of
the vertex operator by the SL(2,C) transformation $z\to 1/z$. This is to be contrasted
with the BPZ transformation in open string theory where we use the SL(2,R) transformation
$z\to -1/z$.
\item Hermitian conjugation reverses the ordering of the oscillators as well
as the relative position of the basis state $\vp_r(k)$ and
the coefficient $\psi_r(k)$, while BPZ
conjugation leaves them unchanged.\footnote{Since BPZ conjugation reverses the
radial ordering, a more correct statement would be
that BPZ conjugation also reverses the order of the operators, but for every 
reordering of a pair of grassmann odd operators, there is a minus sign.}
\end{enumerate}
Therefore if a basis state $|\vp_r\rangle$ has $p_r$ number of $b$, $\bar b$ 
oscillators and $q_r$ number of $c$, $\bar c$ oscillators, the reality condition
of \cite{9206084} may be stated as
\be
\psi_r(k)^* = - \, (-1)^{q_r}\, (-1)^{(p_r+q_r) (p_r+q_r-1)/2} (-1)^{q_r-p_r}
\psi_r(-k)
\ee
where the first minus sign is due to the requirement of relative minus sign between
the hermitian and BPZ conjugation, the second factor is the effect of extra minus signs
picked up by the $c$, $\bar c$ oscillators, the third
factor comes from having to
rearrange the ghost oscillators, and the last factor comes from the 
reversal of the relative position of $\vp_r(k)$ and $\psi_r(k)$. 
Using the relation $n_r=q_r-p_r$, one can see that this
reduces to 
\be
\psi_r(k)^*=(-1)^{n_r(n_r+1)/2+1} (-1)^{n_r}\psi_r(-k)\, .
\ee
This is not quite the same as 
\refb{erealpsi}, but differs from it by a factor of $(-1)^{n_r}
=(-1)^{n_r-2}$. This difference however can be removed by redefining the
reality condition on $\psi_r(k)$ by utilizing the freedom described
in \refb{eamb1} with the choice $\alpha=\pi/2$.

\sectiono{Reality condition in heterotic and type II string field theory} \label{sfermion}

In this section we shall determine the reality condition in the heterotic and type
II string theories. We shall discuss the heterotic string theory in detail and then
briefly mention the results for type II string theory.

\newcommand{\g}{\bar L^G}

The world-sheet theory of heterotic string in ten dimensions 
has additional fields besides what we have in the bosonic string theory.
They include ten
right moving 
fermions $\psi^\mu$, bosonic ghosts
$\beta$, $\gamma$ and an anti-chiral CFT of central charge 16, describing 
either $E_8\times E_8$ or SO(32) current algebra. 
The singular parts of the additional operator product expansions are:
\be 
\psi^\mu(z) \psi^\nu(w) = -{1\over 2(z-w)}\, \eta^{\mu\nu}, \quad
\beta(z) \gamma(w) = {1\over z-w} +\cdots\, . 
\ee
The operator product of $\beta$ and $\gamma$ has non-standard sign
convention, but this is the one that is compatible with the bosonization rules
\refb{eboserule} and the operator product expansion \refb{eboseope} if we
take into account the fact that $\xi,\eta$ anti-commute with $e^{\pm \phi}$.
Alternatively we could include an extra minus sign in the $\beta$-$\gamma$ operator
product expansion and include an extra minus sign in one of the terms in
\refb{eboserule}.

$\psi^\mu$, $\beta$ and $\gamma$ have mode expansions:
\be 
\psi^\mu(z) =\sum_n \psi^\mu_{n} z^{-n-1/2}, \quad \beta(z) =\sum_n
\beta_n z^{-n-3/2}, \quad \gamma(z) = \sum_n \gamma_n z^{-n+1/2}\, . 
\ee
For the anti-chiral CFT of central charge 16,
we shall not use any explicit representation, but 
denote by $|K\rangle$ a basis of Virasoro primary states satisfying
\be \label{ecftg}
\langle K | L\rangle =\delta_{KL}, \quad \langle K | J(1) | L\rangle = \hbox{real}\, .
\ee
If we were representing the theory by a set of left-moving scalars $Y^I$ then
examples of such primary operators would have been $\p Y^I$, $i\cos Y^I$,
$i\sin Y^I$ etc. The full set of states in this CFT are obtained by acting on these
primary states the Virasoro generators $\g_{-n}$ of this CFT. From now
on we shall refer to this
CFT as $CFT_G$, and the anti-holomorphic stress tensor of this CFT by
$\bar T^G$.

For computing string amplitudes, we need to bosonize the $\beta$-$\gamma$
system using the relations
\be \label{eboserule}
\gamma = \eta \, e^\phi, \quad \beta = \p\xi e^{-\phi}\, .
\ee
The leading terms in the operator product of $\xi$, $\eta$ and $e^{q\phi}$ are
\be \label{eboseope}
\xi(z) \eta(w)\simeq {1\over z-w}+\cdots, \quad e^{q\phi(z)} e^{q'\phi(w)} \simeq
(z-w)^{-qq'} e^{(q+q')\phi(w)} + \cdots\, .
\ee 
The fields $\psi^\mu$, $\beta$, $\gamma$ carry odd GSO parity whereas $e^{q\phi}$
carries GSO parity $(-1)^q$ for integer $q$. 
It follows from this that $\xi$ and $\eta$ have even GSO parity.
We also assign $e^{q\phi}$ to have picture number $q$ and ghost number 0, 
$\xi$ to have picture number
$1$ and ghost number $-1$ and
$\eta$ to have picture number $-1$ and ghost number 1,
so that $\beta$ and $\gamma$ have
zero picture number, and ghost numbers $-1$ and $1$ respectively.

The string field has two components: $|\Psi\rangle$ and $|\wt\Psi\rangle$.
If we denote by $\HH_n$ the Hilbert space of GSO even states in string theory
satisfying \refb{el0} and carrying picture number $n$, then $|\Psi\rangle$ takes
value in $\HH_{-1}\oplus \HH_{-1/2}$ and $|\wt\Psi\rangle$ takes value in
$\HH_{-1}\oplus \HH_{-3/2}$. We shall denote by $\HH_{NS}=\HH_{-1}$ the Hilbert space of
NS sector states and by $\HH_R=\HH_{-1/2}\oplus \HH_{-3/2}$ 
the Hilbert space of R sector states.
The string field theory action takes the form
\be \label{eactpsi}
S = g_s^{-2}\left[
-{1\over 2} \langle\wt\Psi |c_0^- Q_B \GG |\wt\Psi\rangle 
+ \langle\wt\Psi |c_0^- Q_B |\Psi\rangle + 
\sum_{n=1}^\infty {1\over n!} \cL \Psi^n\cR
\right]\, ,
\ee
where
\be \label{edefgg}
\GG|s\rangle =\begin{cases} {|s\rangle \quad \hbox{if $|s\rangle\in \HH_{NS}$}\cr
\XX_0\, |s\rangle \quad \hbox{if $|s\rangle\in \HH_R$}}
\end{cases}\, ,
\ee
\be \label{edefchi0}
\XX_0 \equiv \ointop {dz\over z} \XX(z)\, ,
\ee
and $\XX(z)$ is the
picture changing
operator (PCO) given by
\be \label{epicture}
\XX(z) = \{Q_B, \xi(z)\} = c \partial \xi + 
e^\phi T_F - {1\over 4} \p\eta e^{2\phi} b
- {1\over 4} \p\left(\eta e^{2\phi} b\right)\, ,
\ee
where 
\be
T_F(z) = - \psi_\mu \p X^\mu\, .
\ee
The $\ointop$ in \refb{edefchi0} includes the $1/2\pi i$ factor so that
$\ointop dz/z$ is normalized to 1.
The definition of $\cL\Psi^n\cR$ is similar to that for the bosonic string theory with the
following important differences.
Now $\cL A_1\cdots A_m \wh A_1\cdots \wh A_n\cR$ for $m$ NS-sector vertex 
operators $A_1,\cdots A_m$ and  $n$ R-sector vertex operators 
$\wh A_1,\cdots \wh A_n$ 
has, besides the insertion of the vertex operators and the $b$-ghost insertions, also
insertion of PCO's. The locations of the PCO's
appear as extra data in the definition of off-shell amplitudes, and so $\wh\PP_{g,n}$
now has to be replaced by $\wt\PP_{g,m,n}$ whose base is the moduli space
of ordinary Riemann surfaces with $m+n$ punctures together with the 
information on spin structure, and whose fiber, for a genus
$g$ amplitude, contains data on the choice of local coordinates around
the punctures, as well as the locations of $2g-2+m+n/2$ PCO's. 
Given $\Sigma_{g,m,n}\in \wt\PP_{g,m,n}$, we define
$\Sigma^*_{g,m,n}\in\wt\PP_{g,m,n}$ as the Riemann surface whose transition
functions and local coordinates are complex conjugates of those of
$\Sigma_{g,m,n}$, and for which the PCO locations are also complex conjugates
of those on $\Sigma_{g,m,n}$. $\oR_{g,n}$ appearing in \refb{edefcurly} 
now has to be replaced by 
$\oR_{g,m,n}$ -- a (generalized) section of $\wt\PP_{g,m,n}$. Detailed 
procedure for choosing this section avoiding spurious poles can be found in
\cite{1504.00609}. 
We shall impose the additional restriction on $\oR_{g,m,n}$ that it is
invariant under conjugation, i.e. if $\Sigma_{g,m,n}\in\oR_{g,m,n}$ then
$\Sigma^*_{g,m,n}\in\oR_{g,m,n}$.
A GSO even 
basis state and the string field component multiplying it are 
taken to be grassmann
even for even ghost number states in the NS sector and odd ghost number states
in the R-sector, and grassmann odd for odd ghost number states in the NS sector
and even ghost number states in the R sector. For GSO odd basis states the
grassmann parities are taken to be opposite. Even though the string field is
always GSO even, the information on the grassmann parity of GSO odd states is
sometimes useful during intermediate stages of manipulation, e.g. $e^{-\phi}$
will be taken to anti-commute with $\psi^\mu$.

In the NS sector we construct
the basis of states $|\vp_r(k)\rangle$  by acting on the tensor product of the
$-1$ picture
vacuum $e^{-\phi}(0)e^{ik\cdot X}(0)|0\rangle$ 
with momentum $k$ and some primary state $|K\rangle$ 
of $CFT_G$,  by the oscillators
of $b$, $c$, $\bar b$, $\bar c$, $\p X$, $\bar \p X$, $\beta$, $\gamma$ and  
$\bar T^G$,
carrying a net GSO parity of $-1$. We do not allow any extra 
factor of $i$ in the definition
of the basis states except for the factors of $i$ accompanying the
factors of $k^\mu$.
The vertex operators for these states can be built from linear combinations
of GSO even products of (derivatives of)
$\partial X^\mu$, $\bar\partial X^\mu$, $\psi^\mu$, $e^{ik\cdot X}$,
$b$, $c$, $\bar b$, $\bar c$,
$e^{q\phi}$, $\p\phi$, $\p\xi$, $\eta$, $\bar T^G$ and $K$,
 without any explicit factor of 
$i$.
It follows from the operator product expansions of the elementary fields,
and \refb{ecftg}, that the
operator products of the $\vp_r$'s, when expressed in terms of $\vp_{r'}$'s, do not
involve any factors of $i$, except for the factor of $i$ accompanying each factor
of $k^\mu$.

Construction of the vertex operators in the Ramond sector also requires introduction
of spin fields. The spin fields are of two types: chiral fields $S_\alpha$ and anti-chiral
fields $S^\alpha$. The mutually local GSO even 
combinations of spin fields in the matter and
ghost sector are
\be 
e^{-(4n+1)\phi/2} S_\alpha, \quad e^{-(4n-1)\phi/2} S^{\alpha}, 
\ee
and their derivatives and products with the NS sector GSO even operators.
The operator products of these spin fields with each other and the GSO even 
NS sector vertex operators ({\it e.g.} $e^{-(2n+1)\phi}\psi^\mu$)
can be computed from the
following basic operator product expansions:
\ben \label{emainope1}
&& \psi^\mu(z) \, e^{-\phi/2} S_\alpha(w)={i\over 2} (z-w)^{-1/2}
(\gamma^\mu)_{\alpha\beta} e^{-\phi/2}S^\beta(w)+\cdots, \nonumber \\ &&
\psi^\mu(z) \, e^{-\phi/2} S^\alpha(w)={i\over 2} (z-w)^{-1/2}
\gamma^{\mu\alpha\beta} e^{-\phi/2}S_\beta(w)
+\cdots,  \nonumber \\ &&
e^{-\phi/2} S_\alpha(z) \, \,  e^{-3\phi/2}S^\beta(w) 
= \delta_{\alpha}^{~\beta} (z-w)^{-2} e^{-2\phi}(w)+\cdots \, , 
\een  
where
$\gamma^\mu$ are ten dimensional $\gamma$-matrices, normalized as
\be
\{\gamma^\mu, \gamma^\nu\} = 2 \eta^{\mu\nu} \, {\bf 1}\, ,
\ee
where $(\gamma^\mu \, \gamma^\nu)_\alpha^{~\beta} 
\equiv \gamma^\mu_{\alpha\delta} \gamma^{\nu\delta\beta}$ etc.
We shall use a representation in which all the $\gamma$-matrices
are purely imaginary and symmetric:\footnote{Note that the overall
phase of $\gamma^\mu$ can be changed
by phase rotating $S_\alpha$ and
$S^\alpha$ in the opposite direction without affecting
the last equation
in \refb{emainope1}.
Therefore the choice of $\gamma^\mu$ to
be imaginary fixes the phases of $S_\alpha$ and $S^\alpha$. The symmetry 
of $\gamma^\mu$ follows from the consistency of the operator product expansion
\refb{emainope1}. For example evaluation of the three point function
$\langle e^{-\phi} \psi^\mu(z) \, e^{-\phi/2} S_\alpha(w) e^{-\phi/2} S_\beta(y)\rangle$
using \refb{emainope1} in different ways leads to the symmetry of
$\gamma^\mu_{\alpha\beta}$.}
\be \label{sogamma}
(\gamma^\mu_{\alpha\beta})^*=-\gamma^\mu_{\alpha\beta},
\quad (\gamma^{\mu\alpha\beta})^* =- \gamma^{\mu\alpha\beta}, \quad
\gamma^\mu_{\alpha\beta}=
\gamma^\mu_{\beta\alpha}, \quad 
\gamma^{\mu\alpha\beta}=
\gamma^{\mu\beta\alpha}
\, .
\ee
With this the right hand sides of \refb{emainope1} have real coefficients.
If $\Gamma^i$ for $1\le i\le 8$ are the real $8\times 8$
SO(8) gamma matrices
satisfying $\Gamma^i(\Gamma^j)^T + \Gamma^j (\Gamma^i)^T
=2\, \delta^{ij}$ then a specific choice of
SO(9,1) gamma matrices satisfying \refb{sogamma} is given by
\ben
&& \gamma^i_{\alpha\beta} = \pmatrix
{0 & i\, \Gamma^i\cr i\, (\Gamma^i)^T & 0}_{\alpha\beta},
\quad \gamma^{i\alpha\beta}=\pmatrix{0 & -i\, \Gamma^i\cr 
-i\, (\Gamma^i)^T & 0}_{\alpha\beta}
\quad \hbox{for \, $1\le i\le 8$}, 
\nonumber \\
&& \gamma^9_{\alpha\beta}= \pmatrix{i\, I & 0\cr 0 &-i\, I}_{\alpha\beta}, \quad 
\gamma^{9\alpha\beta} = \pmatrix{-i\, I &0\cr 0 & i\, I}_{\alpha\beta}, \quad
\gamma^0_{\alpha\beta} = \gamma^{0\alpha\beta}=
\pmatrix{i\, I & 0\cr 0&  i \, I}_{\alpha\beta}\, .\nonumber\\
\een

From these, and the grassmann parities of various
operators described earlier,
we can derive all other operator products, {\it e.g.} we have the
following useful relations involving GSO even operators:
\ben \label{emainopeagain1}
&& e^{-\phi}\psi^\mu(z) \, e^{-\phi/2} S_\alpha(w)={i\over 2} (z-w)^{-1}
(\gamma^\mu)_{\alpha\beta} e^{-3\phi/2}S^\beta(w)+\cdots, \nonumber \\ &&
e^{\phi}\psi^\mu(z) \, e^{-3\phi/2} S^\alpha(w)=-{i\over 2} (z-w)
\gamma^{\mu\alpha\beta} e^{-\phi/2} S_\beta(w)+\cdots
\, , \nonumber \\
&& e^{-\phi/2} S_\alpha(z)  e^{-\phi/2} S_\beta(w) 
= -i (z-w)^{-1} \gamma^\mu_{\alpha\beta} e^{-\phi(w)} \psi_\mu(w) +\cdots\, ,
\nonumber \\
&& e^{-\phi/2} S_\alpha(z)  e^{\phi/2} S^\beta(w) 
= -\delta_{\alpha}^{~\beta} (z-w)^{-1} +\cdots\, , 
\een
etc. 
We shall choose the basis of $-1/2$ picture states $|\wh\vp_s(k)\rangle$ 
in the R-sector to be such that
their vertex operators are constructed from products of (derivatives of) the
operators appearing on the left hand side of the above equations, and 
other GSO even operators that were used to construct vertex
operators for the basis states in the NS sector,
without any explicit factor of $i$. A similar procedure is followed for
the construction of the GSO even basis states $|\wt\vp_s\rangle$ of
$\HH_{-3/2}$.
In this case all the coefficients appearing in the operator
product expansion of operators representing GSO even basis states
in the NS and R sectors are manifestly
real except for the factor of $i$ multiplying each factor of $k^\mu$.

During the evaluation of superstring amplitudes we also need insertion of 
PCO's given in \refb{epicture}.
Again the operator product of these operators with each other and the
NS and R sector vertex operators do not contain any explicit factors of $i$
except those accompanying the $k^\mu$ factors.

To summarize, we have argued that
as in the case of bosonic string theory, the operator product expansion of the vertex 
operators of basis states in the NS or R sectors, and the PCO's,
do not contain any explicit factor of $i$, except that every factor of
$k^\mu$ is accompanied by a factor of $i$. Using this one can argue, as
in the case of bosonic string theory, that 
\ben \label{ereal1}
&& \cL\vp_{r_1}(k_1)\cdots \vp_{r_m}(k_m) 
\wh\vp_{s_1}(\ell_1)\cdots \wh\vp_{s_n}(\ell_n)
\cR^* \nonumber \\
&=& \cL\vp_{r_1}(-k_1)\cdots \vp_{r_m}(-k_m) 
\wh\vp_{s_1}(-\ell_1)\cdots \wh\vp_{s_n}(-\ell_n)
\cR \, ,
\een
where $\vp_{r_i}$'s denote basis of NS sector vertex operators of picture number $-1$ and
$\wh\vp_{s_i}$'s denote basis of R-sector vertex operators of picture number $-1/2$.
This assumes that we have chosen the integration slices $\oR_{g,m,n}$ of
$\wt\PP_{g,m,n}$ such that it is invariant under conjugation, including locations of the
PCO's.

We can now begin discussing the reality of superstring field theory action.
We begin by stating the reality condition on the field $|\Psi\rangle$.
We expand the NS sector component of $|\Psi\rangle$ as
\be 
|\Psi_{NS}\rangle = \sum_r  \int {d^{{10}} k \over (2\pi)^{10}}
\, \psi_r(k) |\vp_r(k)\rangle
\ee
and the R component of $|\Psi\rangle$ as
\be
|\Psi_R\rangle = \sum_s \int {d^{{10}} k \over (2\pi)^{10}}\, \wh\psi_s(k) |\wh\vp_s(k)
\rangle\, .
\ee
We impose the reality condition
\be \label{e2.16}
\psi_r(k)^* = (-1)^{n_r (n_r+1) /2 + 1}\psi_r(-k)\, ,
\ee
and
\be \label{e2.17}
\wh\psi_s(k)^* = -i \, (-1)^{(n_s+1) (n_s+2) /2}\wh\psi_s(-k)\, ,
\ee
where $n_r$ and $n_s$ are ghost numbers
of $\vp_r$ and $\wh \vp_s$ respectively. As will be seen in \refb{esym1}, the difference
in the exponent of $(-1)$ in \refb{e2.16} and \refb{e2.17} is due to the fact that in
the R sector the grassmann parity of $\wh \vp_s$ is given by $(-1)^{n_s+1}$. 
Defining $\chi_r$, $\wh\chi_s$ via\footnote{We shall use the representation
$i^{1/2}=\exp[i\pi/4]$.}
\be \label{ephase2}
\psi_r(k) = i^{n_r (n_r+1) /2 + 1} \chi_r(k), \quad
\wh\psi_s(k) = i^ {(n_s+1) (n_s+2) /2 + 1/2} \wh\chi_s(k)\, ,
\ee
the reality condition takes the form
\be\label{erchi}
\chi_r(k)^*=\chi_r(-k), \quad
\wh\chi_s(k)^*=\wh\chi_s(-k)\, .
\ee
As in the case of bosonic string theory, we could absorb the phase factors
on the right hand sides of \refb{ephase2} into the definition of the basis states
$|\vp_r(k)\rangle$ and $|\wh\vp_s(k)\rangle$.
In that case $\chi_r(k)$ and $\wh\chi_s(k)$ will be directly interpreted as the
coefficients of expansion of the string field in this basis. However we shall
proceed with the original choice of basis.

Using the reality condition on $|\Psi\rangle$, we can proceed to check
the reality of the interaction term involving $\cL\Psi^n\cR$.
We write this as
\ben \label{eactionNSR}
\sum_n {1\over n!}\cL\Psi^n\cR &=& \sum_{m,n} {1\over m! \, n!} 
\sum_{r_1,\cdots r_m}\sum_{s_1,\cdots s_n} (-1)^{\sum_{i<j} n_{r_i} n_{r_j}
+\sum_{i<j} (n_{s_i}+1)(n_{s_j}+1) + \sum_{i,j} n_{r_i} (n_{s_j}+1)}
\nonumber \\ &&
\int {d^{10} k_1\over (2\pi)^{10}}\cdots {d^{10} k_m \over 
(2\pi)^{10}} {d^{10}\ell_1\over (2\pi)^{10}}
\cdots {d^{10}\ell_n \over (2\pi)^{10}} \nonumber \\ &&
V^{(m,n)}_{r_1\cdots r_m, s_1,\cdots s_n}(k_1,\cdots k_m,\ell_1,\cdots 
\ell_n) \,
\chi_{r_1}(k_1)\cdots \chi_{r_m}(k_m) \wh \chi_{s_1}(\ell_1) 
\cdots \wh \chi_{s_n}(\ell_n)
\, , \nonumber \\
\een
where
\ben\label{edefv}
&& V^{(m,n)}_{r_1,\cdots r_m, s_1,\cdots s_n}(k_1,\cdots k_m, \ell_1,\cdots \ell_n)
\nonumber \\
&=& i^{\sum_{i=1}^m \{n_{r_i} (n_{r_i}+1)/2 + 1\} + \sum_{j=1}^n \{(n_{s_j}+1) (n_{s_j}+2) / 2 + 1/2\}}
\cL\vp_{r_1}(k_1)\cdots \vp_{r_m}(k_m) \wh\vp_{s_1}(\ell_1)\cdots \wh\vp_{s_n}(\ell_n)
\cR \, . \nonumber \\
\een
The sign factors appearing in the first line of \refb{eactionNSR} arise from
having to move the coefficients $\chi_r$ and $\wh\chi_s$ through the
operators $\vp_r$ and $\wh\vp_s$.
We shall define the $V^{(m,n)}$'s for other ordering of the indices and
arguments by appropriately rearranging the order of
$\vp_r$'s and $\wh\vp_s$'s inside $\cL ~\cR$ in \refb{edefv}
using the known grassmann
parity of the basis states. Since the string field is always grassmann even,
with this definition we also have
\ben \label{eactionNSRnew}
\sum_n {1\over n!}\cL\Psi^n\cR &=& \sum_{m,n} {1\over m! \, n!} 
\sum_{r_1,\cdots r_m}\sum_{s_1,\cdots s_n} (-1)^{\sum_{i<j} n_{r_i} n_{r_j}
+\sum_{i<j} (n_{s_i}+1)(n_{s_j}+1) +\sum_{i,j} n_{r_i} (n_{s_j}+1)}
\nonumber \\ &&
\int {d^{10} k_1\over (2\pi)^{10}}\cdots {d^{10} k_m \over 
(2\pi)^{10}} {d^{10}\ell_1\over (2\pi)^{10}}
\cdots {d^{10}\ell_n \over (2\pi)^{10}} \nonumber \\ &&
V^{(m,n)}_{s_n\cdots s_1, r_m,\cdots r_1}(\ell_n,\cdots 
\ell_1,k_m,\cdots k_1) 
\wh \chi_{s_n}(\ell_n)\cdots \wh \chi_{s_1}(\ell_1) \chi_{r_m}(k_m) 
\cdots \chi_{r_1}(k_1)
\, . \nonumber \\
\een
$V^{(m,n)}$ has the symmetry properties
\ben \label{esym1}
&& V^{(m,n)}_{\cdots r_i r_j\cdots} (\cdots, k_i, k_j,\cdots)
=  (-1)^{n_{r_i} n_{r_j}} \, V^{(m,n)}_{\cdots r_j r_i\cdots} (\cdots, k_j, k_i,\cdots) \, ,
\nonumber \\ &&
V^{(m,n)}_{\cdots s_i s_j\cdots} (\cdots, \ell_i, \ell_j,\cdots)
=  (-1)^{(n_{s_i}+1) (n_{s_j}+1)} \, V^{(m,n)}_{\cdots s_j s_i\cdots} (\cdots, \ell_j, \ell_i,\cdots) \, ,
\nonumber \\ &&
V^{(m,n)}_{\cdots r_i s_j\cdots} (\cdots, k_i, \ell_j,\cdots)
=  (-1)^{n_{r_i} (n_{s_j}+1)} \, V^{(m,n)}_{\cdots s_j r_i\cdots} (\cdots, \ell_j, k_i,\cdots) \, ,
\een
where we have used that in the NS sector the grassmann parity is
$(-1)^{n_r}$ whereas in the R sector the grassmann parity is $(-1)^{n_s+1}$.
It follows from \refb{ereal1}, \refb{edefv},
\refb{esym1}, and that the number $n$ of Ramond sector
states is always even, that
\ben \label{evv*}
&& \left(V^{(m,n)}_{r_1,\cdots r_m, s_1,\cdots s_n}(k_1,\cdots k_m, \ell_1,\cdots \ell_n)
\right)^*\nonumber \\
&=& (-1)^{\sum_{i=1}^m \{n_{r_i} (n_{r_i}+1)/2 + 1\} + \sum_{j=1}^n 
\{(n_{s_j}+1) (n_{s_j}+2) / 2 + 1/2\}}
V^{(m,n)}_{r_1,\cdots r_m, s_1,\cdots s_n}(-k_1,\cdots -k_m, -\ell_1,\cdots -\ell_n)
\nonumber \\ 
&=& (-1)^{\sum_{i=1}^m \{n_{r_i} (n_{r_i}+1)/2 + 1\} 
+ \sum_{j=1}^n \{(n_{s_j}+1) (n_{s_j}+2) / 2 + 1/2\}}
\nonumber \\
&& (-1)^{\sum_{i<j} n_{r_i} n_{r_j} + \sum_{i<j} (n_{s_i}+1) (n_{s_j}+1) 
+ \sum_{i,j} n_{r_i} (n_{s_j}+1)} \nonumber \\ &&
V^{(m,n)}_{s_n,\cdots s_1, r_m,\cdots r_1}(-\ell_n,\cdots -\ell_1, -k_m,\cdots -k_1)\, .
\een
Let us define
\be\label{emn1}
M = \sum_{i=1}^m n_{r_i}, \quad N = \sum_{j=1}^n n_{s_j}\, .
\ee
Ghost charge conservation gives
\be\label{emn2}
M+N = 2 (m+n)\, .
\ee
Using \refb{emn1}, \refb{emn2}, 
the exponents in the third and fourth lines of \refb{evv*}
may be written as, respectively,
\ben \label{ex1}
&& \sum_{i=1}^m \{n_{r_i} (n_{r_i}+1)/2 + 1\}
+ \sum_{j=1}^n \{(n_{s_j}+1) (n_{s_j}+2) / 2 + 1/2\} \nonumber \\
&=& {1\over 2} \sum_{i=1}^m n_{r_i}^2 + {1\over 2} \sum_{j=1}^n (n_{s_j}
+1)^2 + {1\over 2} M + m + {1\over 2} N + n \nonumber \\
&=& {1\over 2} \sum_{i=1}^m n_{r_i}^2 + {1\over 2} \sum_{j=1}^n (n_{s_j}
+1)^2 + 2(m+n)\, ,
\een
and
\ben \label{ex2}
&& \sum_{i<j} n_{r_i} n_{r_j} + \sum_{i<j} (n_{s_i} +1) (n_{s_j}+1) 
+
\sum_{i,j} n_{r_i} (n_{s_j}+1) \nonumber \\
&=& {1\over 2} \sum_{i,j} n_{r_i} n_{r_j} + {1\over 2} \sum_{i,j} (n_{s_i} +1) (n_{s_j}+1)
+
\sum_{i,j} n_{r_i} (n_{s_j}  +1)
- {1\over 2} \sum_i n_{r_i}^2 - {1\over 2} \sum_j (n_{s_j}+1)^2
\nonumber \\
&=& {1\over 2} M^2 + {1\over 2} (N+n)^2 + M (N+n) - {1\over 2} \sum_i n_{r_i}^2 - {1\over 2} \sum_j (n_{s_j}+1)^2
\nonumber \\
&=& {1\over 2} (M+N+n)^2 - {1\over 2} \sum_i n_{r_i}^2 
- {1\over 2} \sum_j (n_{s_j}+1)^2
\nonumber \\
&=& {1\over 2} (2m+3n)^2 - {1\over 2} \sum_i n_{r_i}^2 
- {1\over 2} \sum_j (n_{s_j}+1)^2\, .
\een
Using \refb{ex1}, \refb{ex2} and the fact that $n$ is even,
we can express \refb{evv*} as
\be 
\left(V^{(m,n)}_{r_1,\cdots r_m, s_1,\cdots s_n}(k_1,\cdots k_m, \ell_1,\cdots \ell_n)
\right)^* = V^{(m,n)}_{s_n,\cdots s_1, r_m,\cdots r_1}(-\ell_n,\cdots -\ell_1, -k_m,\cdots -k_1)
\, .
\ee
Substituting this into \refb{eactionNSR} and using \refb{erchi}
and \refb{eactionNSRnew} we see that
this is exactly the relation needed for the reality of the interaction term of the
string field theory action.

Let us now turn to the kinetic terms. For this we need to impose reality conditions
on $|\wt\Psi\rangle$ as well. 
We introduce basis states $|\wt\vp_r\rangle$ in $\HH_{-3/2}$ following procedure
similar to that in $\HH_{-1/2}$, expand
$|\wt\Psi\rangle$ as
\be
|\wt\Psi\rangle = \sum_r \int {d^{10} k \over (2\pi)^{10}}\, 
\xi_r(k) |\vp_r(k)\rangle
+ \sum_s  \int {d^{10} k \over (2\pi)^{10}}
\, \wh\xi_s(k) |\wt\vp_s(k)\rangle\, ,
\ee
and impose the reality conditions
\be \label{e2.30}
\xi_r(k)^* = (-1)^{n_r (n_r+1) /2 + 1} \xi_r(-k)\, ,
\quad
\wh\xi_s (k)^* = -i \, (-1)^{(n_s+1) (n_s+2) /2 + 1}\wh\xi_s(-k)\, .
\ee
It is now easy to verify that each of the quadratic terms in the action satisfies
the reality condition. 
Consider for example the term involving fields in $\HH_{-3/2}$:
\be \label{equad}
{1\over 2} \langle \wt\Psi|c_0^- Q_B \GG | \wt\Psi\rangle \nonumber \\
= {1\over 2} \sum_{s_1,s_2} 
\int {d^{10} k_1\over (2\pi)^{10}} {d^{10} k_2\over (2\pi)^{10}} \, 
f_{s_1s_2}(k_1, k_2)\, 
\wh\xi_{s_1}(k_1) \, \wh \xi_{s_2}(k_2)+\cdots\, ,
\ee
where
\be \label{edeff}
 f_{s_1s_2}(k_1, k_2) \equiv (-1)^{(n_{s_1}+1)(n_{s_2}+1)}
\langle \wt\vp_{s_1}(k_1)|c_0^- Q_B \GG | \wt\vp_{s_2}(k_2)\rangle\, .
\ee
It follows from \refb{edeff}, and the fact that the correlation 
functions of $\wt\vp_s$'s do not contain any explicit factor of
$i$ except those accompanying factors of $k^\mu$, that
\be
 f_{s_1s_2}(k_1, k_2)^* =  f_{s_1s_2}(-k_1, -k_2)
\, .
 \ee
 Therefore \refb{equad} gives, using \refb{e2.30}
 \ben \label{equadd}
{1\over 2} \langle \wt\Psi|c_0^- Q_B \GG | \wt\Psi\rangle^* 
&=& {1\over 2} \sum_{s_1,s_2} \,  i^2 (-1)^{(n_{s_1}+1) (n_{s_1}+2) /2 + 1
+ (n_{s_2}+1) (n_{s_2}+2) /2 + 1} \nonumber \\ &&
\int {d^{10} k_1\over (2\pi)^{10}} {d^{10} k_2\over (2\pi)^{10}}
 \,  f_{s_1s_2}(-k_1, -k_2)\, 
 \wh \xi_{s_2}(-k_2) \, \wh\xi_{s_1}(-k_1) \, . \nonumber \\
&=& {1\over 2} \sum_{s_1,s_2} \,  i^2 (-1)^{(n_{s_1}+1) (n_{s_1}+2) /2 + 1
+ (n_{s_2}+1) (n_{s_2}+2) /2 + 1} (-1)^{(n_{s_1}+1) (n_{s_2}+1)} \nonumber \\ &&
\int {d^{10} k_1\over (2\pi)^{10}} {d^{10} k_2\over (2\pi)^{10}}
 \,  f_{s_1s_2}(-k_1, -k_2)\, 
 \wh\xi_{s_1}(-k_1) \, \wh \xi_{s_2}(-k_2) \, . \nonumber \\
\een
Using the ghost charge conservation law $n_{s_2}=4-n_{s_1}$, it is easy to see that the
net pre-factor is unity. Furthermore the signs of $k_i$  in the arguments
of $f_{s_1s_2}$ and $\wh\xi_{s_1}$, $\wh \xi_{s_2}$ 
can be changed by variable redefinition.
Hence we get
\be\label{eampre}
 {1\over 2} \langle \wt\Psi|c_0^- Q_B \GG | \wt\Psi\rangle^*
 =  {1\over 2} \langle \wt\Psi|c_0^- Q_B \GG | \wt\Psi\rangle\, .
 \ee
Similar analysis can be used to establish the
reality of all other quadratic terms  in the action.

As in the case of bosonic string theory, the reality conditions 
\refb{e2.16}, \refb{e2.17} and \refb{e2.30}
are not fixed unambiguously. Besides the ambiguity described in
\refb{eamb1} (with similar phase rotations acting on $\wh\psi_s$,
$\xi_r$ and $\wh \xi_s$) we also have the freedom of multiplying each 
Ramond sector field by an additional factor of
$-1$ under complex conjugation
since the Ramond sector states always occur
in pairs.

The analysis of the reality condition in type II string theory is similar. There are
now four sectors. The action takes the same form as given in 
\refb{eactpsi} with $|\Psi\rangle$ taking value in $\HH_{-1,-1} \oplus
\HH_{-1,-1/2} \oplus \HH_{-1/2, -1}\oplus \HH_{-1/2, -1/2}$, and
$|\wt\Psi\rangle$ taking value in $\HH_{-1,-1} \oplus
\HH_{-1,-3/2} \oplus \HH_{-3/2, -1}\oplus \HH_{-3/2, -3/2}$.  The definition of
$\cL ~\cR$ now includes insertion of holomorphic and anti-holomorphic
PCO's, and the
operator $\GG$ takes the form
\be \label{edefggii}
\GG|s\rangle =\begin{cases} {|s\rangle \quad \hbox{if $|s\rangle\in \HH_{NSNS}$}\cr
\XX_0\, |s\rangle \quad \hbox{if $|s\rangle\in \HH_{NSR}$}\cr 
\bar\XX_0\, |s\rangle \quad \hbox{if $|s\rangle\in \HH_{RNS}$}\cr 
\XX_0\bar\XX_0\, |s\rangle \quad \hbox{if $|s\rangle\in \HH_{RR}$}\cr 
}\, .
\end{cases}
\ee
Analysis similar to the one for heterotic string field theory can be carried out here.
It is easy to see that the reality condition on the fields is determined simply by whether
the field is fermionic or bosonic, since this determines the relation between the
grassmann parity and the ghost number. Therefore once we have chosen a basis
generated by products of derivatives of $e^{ik\cdot X}$ and standard ghost and
matter fields with real operator product expansion coefficients, and expanded the
string field in such a basis, the reality condition on the string field in the NSNS sector
and RR sector takes the form given in \refb{e2.16} whereas for string fields in the
RNS or NSR sectors, the reality condition takes the form of \refb{e2.17}.

Again, once the reality condition is determined, we can use it to fix the sign of the
action. For the heterotic string theory we can consider the component of the
string field $|\Psi\rangle$
describing a graviton field component $h_{12}(k)$:
\be
\int {d^{10} k\over (2\pi)^{10}} \, h_{12}(k) \, 
\bar c \, c \, e^{-\phi} \, \psi^1 \, \bar \p X^2 \, e^{ik\cdot X}\, .
\ee
In the NS sector we can take $\wt\Psi=\Psi$. Substitution into \refb{eactpsi} yields
the kinetic term
\be 
-{1\over 32 g_s^2} \int {d^{10} k\over (2\pi)^{10}} \,  \left[
k^2 \, h_{12}(k) h_{12}(-k)  
+ \hbox{terms proportional to $k^1$, $k^2$}\right ]\, .
\ee
This has the correct sign and no $g_s^2\to -g_s^2$ substitution is
necessary.

In type II string theory we work in the NSNS sector where again we can set the
string field components $\wt\Psi$ and $\Psi$ to be equal from the beginning.
We can again consider the string field component describing the graviton field
\be 
\int {d^{10} k\over (2\pi)^{10}} \, h_{12}(k) \, 
\bar c \, c \, e^{-\phi} \, \psi^1 \, e^{-\bar\phi} \bar\psi^2  \, e^{ik\cdot X}\, .
\ee 
Substitution into the action \refb{eactpsi} gives
\be
{1\over 32 g_s^2} \int {d^{10} k\over (2\pi)^{10}} \,  \left[
k^2 \, h_{12}(k) h_{12}(-k)  
+ \hbox{terms proportional to $k^1$, $k^2$}\right ]\, .
\ee
This has wrong sign and hence we need to make a $g_s^2\to -g_s^2$
substitution to get the correct sign of the kinetic term.

\sectiono{Reality condition as a relation between hermitian conjugation and BPZ
conjugation} \label{srelation}

We have seen that in the case of bosonic string theory, the reality
condition can be interpreted as the equality between hermitian conjugate
and BPZ conjugate of the string field up to a 
sign. We shall now show that
the same result holds for superstring theory provided we choose the
hermitian conjugation rules of various fields 
appropriately, and exploit the ambiguities mentioned in the paragraph
below \refb{eampre} judiciously.\footnote{I would like to thank
Barton Zwiebach for prompting me to investigate this.}
We shall discuss the case of heterotic
string theory in detail; 
the analysis for type II string theory is very similar and will
be mentioned briefly at the end.

We begin by defining the action of hermitian
conjugation on various oscillators. We choose the following definitions
\ben
&& (i\alpha^\mu_{n})^\dagger = -i\alpha^\mu_{-n}, \quad 
(i\bar\alpha^\mu_{n})^\dagger = -i\bar\alpha^\mu_{-n}, \quad 
b_n^\dagger = b_{-n}, \quad c_n^\dagger = c_{-n}, \quad
\bar b_n^\dagger = \bar b_{-n}, \quad \bar c_n^\dagger = \bar c_{-n}, 
\nonumber \\ &&
(\g_n)^\dagger = \g_{-n}, \quad
(\psi^\mu_{n})^\dagger=\psi^\mu_{-n}, \quad \beta_{n}^\dagger = 
\beta_{-n}, \quad \gamma_n^\dagger = -\gamma_{-n}\, .
\een
It is easy to verify that the hermitian conjugation rules given above
preserve the
(anti-)commutation relations between the oscillators.
Besides this we shall assume that for integer $q$ hermitian conjugation
takes $e^{q\phi}$ to $e^{q\phi}$, and it takes the vacuum 
$|k,K\rangle \equiv e^{ik\cdot X}(0)|0\rangle\otimes |K\rangle$ to
$\langle -k, K|$ where $\langle k,K|$ denotes the BPZ conjugate
of $|k,K\rangle$.
Finally it reverses the order of all the operators
and complex conjugates any multiplicative coefficient.
Action of
BPZ conjugation is standard, except that due to half-integral dimensions
carried by various operators we have to choose the phase appropriately.
For example acting  on a primary operator $V(z,\bar z)$ 
of dimension $(\bar h,h)$ the
BPZ conjugation gives a multiplicative factor of $(-1 /z^2)^{h}
(-1/\bar z^2)^{\bar h}
V(1/z,1/\bar z)$, and
we have to fix the phase for non-integer values of $(h-\bar h)$.
We use the convention that acting on a primary operator $V$ of dimension
$(\bar h, h)$ at $z=\bar z=1$, the BPZ conjugation takes it to
\be \label{ebpz2}
e^{-i\pi(h-\bar h)} V(1)\, .
\ee
We also use the convention of \cite{9705038} to define star conjugation as the 
hermitian conjugation followed by inverse of BPZ conjugation. Our goal will
be to check if the string field, satisfying the reality conditions
\refb{e2.16}, \refb{e2.17}, has simple
properties under star conjugation.

We begin our analysis with the NS sector. Let us consider an arbitrary basis
state obtained by acting on the vacuum $e^{-\phi}(0)|k,K\rangle$ by various 
modes of  $b,\bar b,c,\bar c,\psi^\mu,\beta,\gamma,i\alpha^\mu,
i\bar\alpha^\mu$ and $\g$ without any additional factor of $i$.
Let $n_b$ be the number of
$b,\bar b$ oscillators, $n_c$ be the number of $c,\bar c$ oscillators, 
$n_\psi$ be the number of $\psi^\mu$
oscillators, $n_\beta$ be the number of $\beta$ oscillators and $n_\gamma$ be the
number of $\gamma$
oscillators.
We also define
\be \label{ess1}
n_{bc}=n_c-n_b, \quad n_{\beta\gamma}=n_\gamma-n_\beta, \quad
n = n_{\beta\gamma} + n_{bc}\, .
\ee
$n$ is the total ghost number of the state.
Without loss of generality, we can arrange the oscillators such that all the
$b,c,\bar b,\bar c$ oscillators are to the extreme left, all the $\beta,\gamma$
oscillators are grouped together in the middle 
and all the $\psi^\mu$ oscillators are to the extreme right, sitting next to the
vacuum $e^{-\phi}(0)|k,K\rangle$. Locations of the $i\alpha^\mu_{-n}$,
$i\bar\alpha^\mu_{-n}$ and $\g_{-n}$ oscillators will not matter; we can for
definiteness fix them to be at the left of the $\psi^\mu_{-n}$'s.

We shall now collect various factors that arise from star conjugation. First
of all star conjugation changes the relative position of the $b,c,\bar b,\bar c$
oscillators with respect to
the combination of $\psi^\mu$ oscillators and $e^{-\phi}$.
This gives a factor
\be \label{ess5}
(-1)^{(n_{b}+n_c) (n_\psi+1)}\, .
\ee
Star conjugation of the $\beta,\gamma$ system gives a factor
\be \label{ess2}
e^{i\pi (3n_\beta - n_\gamma)/2} (-1)^{n_\gamma} = e^{-3i\pi n_{\beta\gamma} / 2}\, .
\ee 
Here the first factor on the left hand side
is the inverse of the phase described 
in\refb{ebpz2} picked up during inverse
BPZ conjugation while the second factor is due to the minus sign
picked up by the $\gamma$ oscillators during hermitian conjugation.
The star conjugation of $b,c,\bar b,\bar c$ oscillators gives
\be \label{ess3}
(-1)^{n_c} (-1)^{(n_b+n_c) (n_b+n_c-1)/2} = (-1)^{n_{bc} (n_{bc}+1)/2}\, .
\ee
The first factor again comes from \refb{ebpz2} while the second factor
is due to the reversal of order of the $b,c,\bar b, \bar c$
oscillators due to hermitian conjugation.
The star conjugation of $\psi^\mu$ and $e^{-\phi}$ system generates the factor
\be \label{ess4}
(-1)^{n_\psi (n_\psi+1)/2} e^{i\pi(n_\psi+1)/2}\, .
\ee
The first factor is due to the reversal of the order of the operators
due to hermitian conjugation and the second factor comes from \refb{ebpz2}.
There are
no factors from star conjugation of the $i\alpha^\mu_{-n}$,
$i\bar\alpha^\mu_{-n}$ or $\g_{-n}$ since hermitian and BPZ conjugation act on
them in the same way, and they are all grassmann even.

Now the
condition that the state is GSO even requires $n_\psi+1 - n_{\beta\gamma}$ to be even.
Hence we write
\be\label{ess6}
n_\psi+1 -n_{\beta\gamma} = 2 \, m, \quad m\in \ZZZ\, .
\ee
Using \refb{ess1}, \refb{ess6} 
we can express the product of \refb{ess5}-\refb{ess4} as
\be 
(-1)^{n(n+1)/2}\,  .
\ee
This is the sign picked up by a basis state under star conjugation. 
The only other change is the replacement of $k^\mu$ by $-k^\mu$.
Combining this
with  \refb{e2.16} we see that the phase picked up by $\psi_r$ combines
with that of $\vp_r$ to give a net factor of $-1$. However since star conjugation
exchanges the positions of $\psi_r$ and $\vp_r$, it produces another factor
of $(-1)^{n_r-2}$ since this is the grassmann parity of $\psi_r$ and
$\vp_r$.  
This factor, however, can be removed by modifying the reality condition
on $\psi_r$ using the freedom described
in \refb{eamb1} with the choice $\alpha=\pi/2$. With this,
the reality condition on the NS sector string field may be
written as the statement that the star conjugation changes the sign of the string field.

The above analysis can be extended to 
the Ramond sector with a few changes.
We represent the basis states in the same way, as oscillators acting
on the Ramond 
vacuum state $e^{-\phi/2}S_\alpha(0)|k,K\rangle$. We define the action of
hermitian conjugation on the operator $e^{-\phi/2}S_\alpha$ such that it
differs from BPZ conjugation by a factor of $i$.  This may seem unusual, but is
needed for example to satisfy
\be
\langle b | \OO | a\rangle^* = \langle a_{hc}| \OO^\dagger| b_{hc}\rangle
\ee
with the choice $a=e^{-\phi/2} S_\alpha$, $b=e^{-\phi/2} S_\beta$ and 
$\OO=e^{-\phi} \psi^\mu$. Here the subscript $hc$ denotes hermitian 
conjugation. Using the hermitian conjugation rules for $|a\rangle$ and $|b\rangle$
defined above, the result that $\OO^\dagger=-\OO$ due to the exchange
of $e^{-\phi}$ and $\psi^\mu$ induced by hermitian conjugation, and the fact that 
$\gamma^\mu$'s are imaginary and symmetric, we get both the
left and the right hand sides to be $i\gamma^\mu_{\alpha\beta}/2$. However 
without the factor of $i$ included in the definition of the hermitian conjugation of
$e^{-\phi/2}S_\alpha$, the two sides will differ by a minus sign.

We can now compare the signs picked up by a general basis state under
star conjugation with the corresponding
analysis in the NS sector.
The first difference is the extra factor of $i$ in the hermitian conjugation of the Ramond
vacuum.
The second difference
arises from the fact that the 
phase \refb{ebpz2} picked up by the
operator
$e^{-\phi/2}S_\alpha$ during BPZ conjugation has already been taken into account
in the statement that star conjugation of this gives a factor of $i$; so we do not need
to include this in the analog of \refb{ess4}.
However like $e^{-\phi}$,
the new operator is also grassmann odd, hence the effect of reordering generates the
same factor as in \refb{ess4}.
This has the effect of changing
\refb{ess4} to
\be \label{ess4r}
(-1)^{n_\psi (n_\psi+1)/2} e^{i\pi n_\psi /2}\, .
\ee
Another change occurs 
in \refb{ess6} since the requirement of GSO even state now
requires
$n_\psi - n_{\beta\gamma}$ to be even.
Hence we write
\be\label{ess6r}
n_\psi -n_{\beta\gamma} = 2 \, m, \quad m\in \ZZZ\, .
\ee
Using \refb{ess1}, \refb{ess6r} 
we can now express the product of \refb{ess5}-\refb{ess3}, \refb{ess4r} and $i$
as
\be
-i \, (-1)^{(n+1)(n+2)/2} \, .
\ee
This is the sign picked up by $\wh\vp_s$ under star conjugation.
Combining this with \refb{e2.17}, and the fact that the exchange of the position
of $\wh\psi_s$ and $\wh\vp_s$ under star conjugation gives rise to an additional
factor of $(-1)^{n_s-1}$,
 we see that the reality 
condition on the string field requires
that the Ramond sector string field picks up a factor
of $- (-1)^{n_s-1}$ under star conjugation. 
However we can remove the last $(-1)^{n_s-1}$ factor by a combination of the 
freedom described in \refb{eamb1} with $\alpha=\pi/2$ and the freedom
of multiplying each R sector states by an additional factor of $-1$ under
star conjugation.  With this change of star conjugation rules of the R sector field,
we see that the reality condition on the R sector fields can be stated as the condition
that they change sign under star conjugation.

The analysis for the $\wt\Psi$ field is similar, with $e^{-3\phi/2}S^\alpha$ 
replacing $e^{-\phi/2} S_\alpha$ as the operator creating the
Ramond vacuum state.

Let us now briefly discuss the analysis in the type II string theory. We begin with the
NSNS sector. In this case the vacuum is obtained as $e^{-\phi}(0) e^{-\bar\phi}(0)
|k\rangle$ and we have additional oscillators of $\bar \psi^\mu$. Now if we go back
to the analysis of NS sector of the heterotic string theory, we can see that there was no
difference in our treatment of $\psi^\mu$ and $e^{-\phi}$ since they have identical 
transformation under star conjugation and carry identical grassmann and GSO parity,
and due to this all relations \refb{ess5}-\refb{ess6} 
involved only the combination $n_\psi+1$. So we can now
repeat the analysis by grouping the oscillators of $\psi^\mu$, $\bar\psi^\mu$ and the
$e^{-\phi}$, $e^{-\bar\phi}$ together. If $n_\psi$ is the total number of $\psi^\mu$ and
$\bar\psi^\mu$ oscillators then the analysis of the NS sector for heterotic string theory
can be repeated without any change, except that all factors of $(n_\psi+1)$ will be replaced
by $n_\psi+2$ to take into account the presence of the 
$e^{-\bar\phi}$ factor.\footnote{There is actually an additional
factor of $-1$ for every half-integer weight anti-holomorphic field since,
according to \refb{ebpz2},
under BPZ conjugation a half integer weight anti-holomorphic field picks
an additional $-1$ factor compared to a holomorphic field of the same weight.
However since GSO projection ensures that the total number of half-integer 
weight anti-holomorphic fields for any state is even, this does not introduce
any net factor.}
Since the
reality conditions on the string field components take form identical to that for the NS sector
of the heterotic string theory, we conclude that the string field satisfying the reality
condition changes sign under star conjugation.

The analysis in the RR sector follows in  similar fashion once we note that the
operator $e^{-\phi/2} S_\alpha e^{-\bar\phi/2} \bar S_{\beta}$ that creates the
RR vacuum is invariant under star conjugation. The two factors of $i$ picked up
by the two spin fields cancel against the minus sign that comes from having to
exchange their positions. Therefore the analysis of the heterotic string NS sector
can now be repeated with the replacement of $n_\psi+1$ by $n_\psi$ since we
no longer have the $e^{-\phi}$ factor. The result again is the change in sign of
the string field under star conjugation.

For the RNS and NSR sectors, we can use the analysis used for the R sector of
the heterotic string theory with the replacement of $n_\psi$ by $n_\psi+1$ to take into
account the extra factor of $e^{-\phi}$ or $e^{-\bar\phi}$ coming from the NS sector
on the right or left. Therefore the phase picked up by the basis states 
under star conjugation is identical to that for the R sector of heterotic string theory.
Since the component fields also pick up the same phases as in the R sector of the
heterotic theory, we again conclude that the string field changes sign under
star conjugation.

\sectiono{Non-trivial background} \label{sgen}

Let us now consider the effect of putting string theory in a non-trivial space-time
background.  First let us consider the case of bosonic string theory.
In this case in the world-sheet theory
certain number of $X^\mu$'s will be replaced by an 
internal CFT of the same central charge. 
As long as we can choose a basis of conformal 
primary operators of this CFT that has the
property that the three point functions on the sphere of all the primary operators
are real for real insertion points, 
we can build the basis in $\HH$ by taking the tensor product of descendants
of these basis states and the basis states in the CFT involving the remaining
$X^\mu$'s and the ghost fields, constructed in the manner described in
\S\ref{sboson}. With this choice of basis the reality of the string field 
theory action
follows in a manner identical to that in the flat background.

Note that this choice of basis states will typically make the basis non-eigenstates
of the charge operators. For example for a compact internal dimension $Y$, it will
require us to use the operators $e^{i K\cdot Y}+e^{-iK\cdot Y}$ and
$-i(e^{i K\cdot Y}-e^{-iK\cdot Y})$ as basis states, instead of
$e^{\pm iK\cdot Y}$. However the proof of the reality of the string field theory is 
simplest in this basis.

We also need to ensure that the kinetic terms of the string fields, obtained after
imposing the reality condition, come with the correct choice of sign. This will require
the primary states of the basis, chosen in the manner described above, to have
positive  BPZ inner product, with BPZ inner product as defined in \refb{ebpz}.
In order that a CFT provides a consistent background for formulating string  theory,
its correlation functions must satisfy these conditions.

The analysis for heterotic and type II superstring theories are similar.
For example for the heterotic string theory 
we have to assume that the internal superconformal field theory has
a basis of GSO odd and GSO even primary states in the NS sector, and GSO 
odd
and GSO even primary states in the Ramond sector such that the 3-point
functions of $e^{-2n\phi}$ multiplied by
GSO even states in the NS sector,
$e^{-(2n+1)\phi}$ multiplied by the GSO odd states in the NS sector,
$e^{-(4n+1)\phi/2}$ multiplied by the GSO even states in the R sector and
$e^{-(4n-1)\phi/2}$ multiplied by the GSO odd states in the R-sector
are all real for real insertion points. Once this condition is satisfied, the reality
of the string field theory action follows from the same line of argument as 
in the case of string theory in flat space-time background. We also need to check
that once the reality condition is satisfied, the kinetic terms have the correct
sign.  The requirement of reality of the type II string field theory action is a
straightforward generalization of these constraints. 

\bigskip

\noindent {\bf Acknowledgement:}
We wish to thank Roji Pius and Barton Zwiebach for useful discussions and
Barton Zwiebach for critical
comments on an earlier version of the manuscript.
We also thank the Pauli Center for Theoretical Studies at ETH,
Zurich, SAIFR-ICTP, Sao Paulo, Theoretical Physics group of the University
of Torino and LPTHE, Paris for hospitality
during my visit when part of this work was done.
This work was
supported in part by the 
DAE project 12-R\&D-HRI-5.02-0303 and J. C. Bose fellowship of 
the Department of Science and Technology, India.

\end{document}